\documentstyle[sprocl,psfig]{article}
\def\Journal#1#2#3#4{{#1} {\bf #2}, #3 (#4)}

\def\NPB{{\em Nucl. Phys.} B}
\def\PLB{{\em Phys. Lett.}  B}
\def\PRL{\em Phys. Rev. Lett.}
\def\PRD{{\em Phys. Rev.} D}
\def\ZPC{{\em Z. Phys.} C}
\def\MPLA{{\em Mod. Phys. Lett.} A}

\begin{document}
\thispagestyle{empty}
\begin{titlepage}
\begin{center}
January 16, 1998  \hfill UCB-PTH-98/05  \\
~{} \hfill hep-ph/9801331\\

\renewcommand{\thefootnote}{\fnsymbol{footnote}}
\vskip .1in

{\large \bf Probing Physics at Short Distances\\with Supersymmetry}%
\footnote{This work was supported in part by the U.S.  Department of
  Energy under Contracts DE-AC03-76SF00098, in part by the National
  Science Foundation under grant PHY-95-14797, and also by Alfred P.
  Sloan Foundation.}\footnote{To appear in `Perspectives on
  Supersymmetry', World Scientific, Editor G. Kane.}

\vskip 0.3in

Hitoshi Murayama

\vskip 0.05in

{\em Theoretical Physics Group\\
     Ernest Orlando Lawrence Berkeley National Laboratory\\
     University of California, Berkeley, California 94720}

\vskip 0.05in

and

\vskip 0.05in

{\em Department of Physics\\
     University of California, Berkeley, California 94720}

\end{center}

\vfill

\begin{abstract}
We discuss the prospect of studying physics at short distances, such
as Planck length or GUT scale, using supersymmetry as a probe.
Supersymmetry breaking parameters contain information on all physics
below the scale where they are induced.  We will gain insights into
grand unification (or in some cases string theory) and its symmetry
breaking pattern combining measurements of gauge coupling constants,
gaugino masses and scalar masses.  Once the superparticle masses are
known, it removes the main uncertainty in the analysis of proton
decay, flavor violation and electric dipole moments.  We will be able
to discuss the consequence of flavor physics at short distances
quantitatively.  
\end{abstract}

\vfill

\end{titlepage}

\newpage
\setcounter{page}{1}
\setcounter{footnote}{0}
\renewcommand{\thefootnote}{\alph{footnote}}

\title{Probing Physics at Short Distances with Supersymmetry}
\author{Hitoshi Murayama}
\address{ Department of Physics, University of California,
Berkeley, CA 94720}
\maketitle\abstracts{
We discuss the prospect of studying physics at short distances, such
as Planck length or GUT scale, using supersymmetry as a probe.
Supersymmetry breaking parameters contain information on all physics
below the scale where they are induced.  We will gain insights into
grand unification (or in some cases string theory) and its symmetry
breaking pattern combining measurements of gauge coupling constants,
gaugino masses and scalar masses.  Once the superparticle masses are
known, it removes the main uncertainty in the analysis of proton
decay, flavor violation and electric dipole moments.  We will be able
to discuss the consequence of flavor physics at short distances
quantitatively.  }

\section{Introduction}

The aim of particle physics is very simple: to understand the
structure of matter and their interactions at as short distant scale
as possible.  This is the ultimate form of the reductionist approach
of physics.  This approach has revealed several layers of
distance scales in nature, bulk (1~cm), atomic ($10^{-8}$~cm), nuclear
($10^{-13}$~cm).  Understanding physics at shorter distance scales
always gave us better understanding of physics at a previously known
longer distance scale.  Knowing the structure of atoms, we can deduce
the chemical properties of atoms and molecules.  Knowing the
statistics of nuclei, we understand the levels of molecular
excitations.  The quest continues to understand the origin of the
known distance scales, such as the electroweak scale of $10^{-16}$~cm,
and to discover new layer of physics at shorter distance scales.  

The main motivation of supersymmetry is to stabilize the electroweak
scale against radiative corrections, which tend to make it much
shorter (as short as the Planck length) or much longer (no electroweak
symmetry breaking).  Whenever we speculate about physics at shorter
distance scales, we cannot go around the problem of the stability of
the electroweak scale.  However, once we accept supersymmetry as the
stabilization mechanism, we are allowed to speculate physics at much
shorter distances, and ask questions about the origin of gauge forces,
fermion masses, and even cosmological issues such as baryon
asymmetry.  

The aim of this short article is to further elaborate on this point.
Not only that supersymmetry allows us to speculate physics as the
shortest distance scales, it actually provides probes of it.  One can
even dream about exploring physics at the GUT scale ($10^{-30}$~cm) or
the Planck scale ($10^{-33}$~cm) once we see superparticles.  We will
present various possibilities how we may be able to probe physics at
such short distance scales using supersymmetry as a probe (see, {\it
  e.g.}\/, a review on this point\,\cite{ICEPP}).  We therefore assume
that we will find and can study superparticles at collider
experiments.\footnote{Needless to say, it is important to confirm
  experimentally that the discovered new particles have properties
  consistent with supersymmetry.\,\cite{FPMT,FNT} }

Of course, such a dream scenario cannot be discussed without certain
assumptions.  For each examples of such probes in the following
sections, we try to make explicit what the underlying assumptions are.
One of the main assumptions in any of these discussions is that the
layers in distance scales are exponentially apart from each other.
This is not an unreasonable assumption from the historic perspective.
All the layers of physics came at very disparate distance scales.
There appears to be nothing new between the characteristics scales:
{\it deserts}\/.  We do not know if this is the way nature is
organized; we can only assume that the shorter distance scales also
come with exponential hierarchy and discuss its consequences in our
further exploration of physics at yet shorter distance scales.

\section{Grand Unification}

The simplest example in our approach is the grand unification.
Needless to explain, a grand unified theory intends to explain the
rather baroque pattern of quark, lepton quantum numbers in the
Standard Model by embedding its gauge groups into a simple gauge
group.  Such a theory would resolve bizarre puzzles in the Standard
Model.  The fact that the matter is neutral (at the level of at least
$10^{-21}$) requires a cancellation of electric charges between and
electron, two up-quarks and one down-quark.  The cancellation of
anomalies in the Standard Model appears miraculous.  And probably most
importantly, the grand unification explains why the strong interaction
is stronger than the electromagnetism as a simple consequence of the
difference in the size of the gauge groups.  

\subsection{Gauge Coupling Constants}

The supersymmetric grand unification received a strong attention in
the past seven years after the precise measurement of the weak mixing
angle at LEP in 1991.  Many took the agreement of the observed and
predicted value of the weak mixing angle as an experimental support
for supersymmetry because the prediction was quite off if the minimal
non-supersymmetric Standard Model was used to predict the weak mixing
angle.  The situation has not changed qualitatively since.  The
detailed discussions on the dependence on superparticle spectrum,
GUT-scale threshold, and its correlation to the proton decay are all
important issues, and we refer to another chapter on gauge
unification.

Since our question is what we will learn by using supersymmetry as a
probe, let us suppose that we already have found the superparticles.
Then the question on grand unification changes dramatically.  First of
all, we do not need to motivate supersymmetry {\it assuming}\/ grand
unification.  The question goes the other way around.  Since
we know that the supersymmetry is there, we will rather ask if the
gauge coupling constants unify given the particle content seen at the
electroweak scale.  More importantly, we measure all the masses of
superparticles, which give us quantitative inputs on the supersymmetry
thresholds in the renormalization group analysis.  Knowing the
particle content at the TeV scale and their masses will completely
change the rule of the research.  Note, however, that this analysis
assumes a desert between the electroweak scale under experimental study
and the GUT scale. 

If they do unify within a certain accuracy, say within a few percents,
we will begin asking what the origin of the small mismatch is (if
any).  For each of the GUT models we construct, we calculate the
GUT-scale threshold corrections and compare them to the data.  Such an
analysis would certainly exclude parts of the parameter space in each
model, and in some cases, the model itself.  Especially the
correlation to the proton decay becomes important, since the GUT-scale
threshold corrections contain information about the mass of the
color-triplet Higgsino which mediates the dimension-five proton decay
such as $p\rightarrow e^+ K^0$ and the mass of the GUT gauge bosons
which mediate the dimension-six proton decay such as $p\rightarrow e^+
\pi^0$.\,\cite{HMY} We will come back to this point in the section on
proton decay.  Another origin of a small mismatch may be a higher
dimension operator in the gauge kinetic function which depends on the
GUT-Higgs field, such as $\int d^2 \theta \mbox{Tr}(\frac{\Sigma}{M}
W_\alpha W^\alpha)$, where $\Sigma$ is the adjoint Higgs in $SU(5)$
GUT.\,\cite{Hall-Sarid,Nath}  If we assume that $M$ is the reduced Planck
scale and $\Sigma$ has a VEV at the conventional GUT-scale of order
$10^{16}$~GeV, such an operator gives an order percent correction to
the gauge coupling unification.  This possibility may unfortunately
contaminate the information on GUT-scale threshold.  In any case,
however, it is clear that the rule of the game changes from motivating
supersymmetry using GUT to making selection of GUT models from
observed supersymmetry spectrum.

\begin{figure}[t]
  \centerline{
  \psfig{file=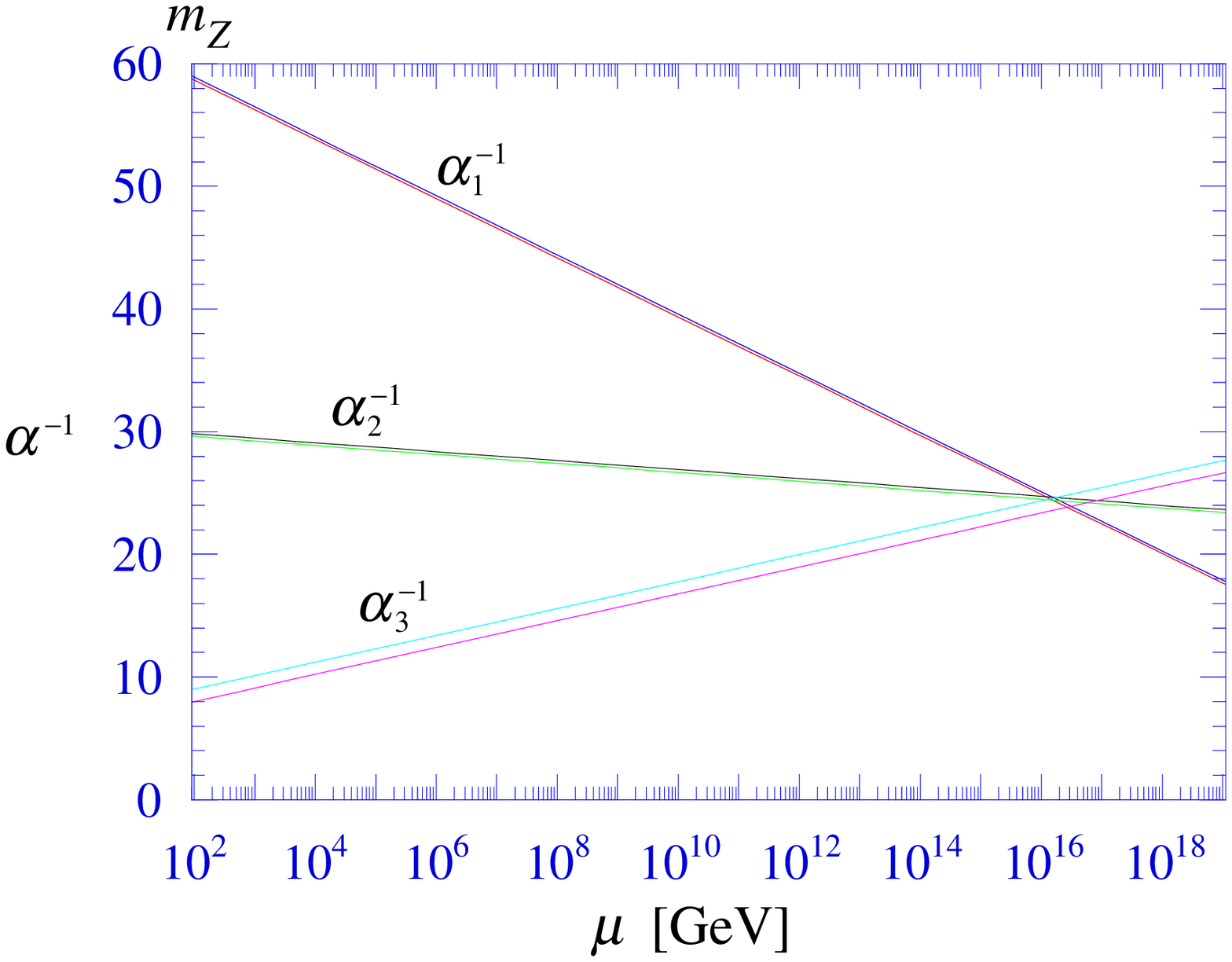,width=0.49\textwidth}
  \psfig{file=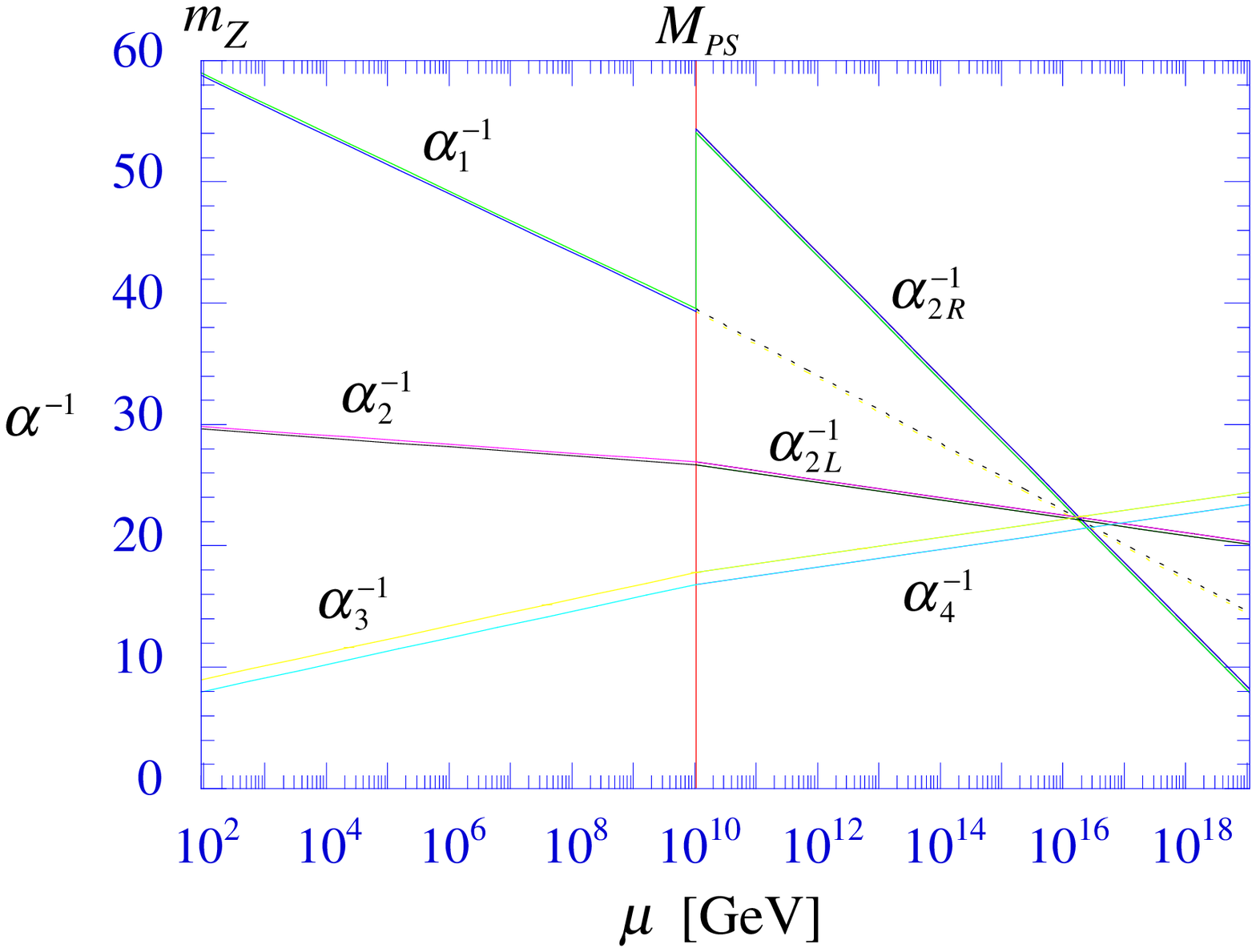,width=0.49\textwidth}
  }
  \caption[coupling]{The renormalization group evolution of the gauge
  coupling constants in two models, $SU(5)$ GUT with grand desert, and
  $SO(10)$ GUT with an intermediate Pati--Salam symmetry and a
  particular particle content above the Pati--Salam scale.\,\cite{KMY1}}
\label{Fig:gauge}
\end{figure}

Unfortunately, the fact that the observed gauge couplings appear
consistent with the $SU(5)$ unification does not rule out other
possibilities, such as intermediate gauge groups with certain matter
content.\,\cite{DKR,KMY1} Fig.~\ref{Fig:gauge} shows two patterns of
gauge coupling unifications, one with grand-desert $SU(5)$ and the
other with intermediate Pati--Salam symmetry.  The latter model is
intended to be a comparison toy model to make the points clear for the
later discussions, how the study of superparticles would help us to
sort out the physics at shorter distance scales.

\subsection{Gaugino Masses}

Another aspect of the grand unification is the unification of
superparticle masses.  This discussion assumes that the supersymmetry
breaking parameters are generated at a scale higher than the
GUT-scale, such as the string or Planck scales, and hence respect
grand-unified symmetry.  Under this assumption, we will see if the
superparticle masses unify {\it at the same scale}\/ as where the
gauge coupling constants unify.  In fact, the gaugino mass
unification,
\begin{equation}
  \frac{M_1}{\alpha_1} = \frac{M_2}{\alpha_2} = \frac{M_3}{\alpha_3}
\end{equation}
holds even when the GUT-group breaks to the Standard Model gauge group
in several steps, {\it i.e.}\/, with intermediate gauge groups such as
Pati--Salam $SU(4)\times SU(2) \times SU(2)$ or its subgroup starting
from $SO(10)$ or $E_6$, as long as the Standard Model gauge groups are
embedded in a simple group with a single gaugino mass.\,\cite{KMY1} At
low-energy, the gaugino masses run in the exactly the same way as the
gauge coupling constants squared do, which can be read off from
Fig.~\ref{Fig:gauge} in these two examples.  Therefore, the gaugino
mass unification tests the idea of grand unification in a highly
model-independent manner.

Experimental strategies have been discussed how to disentangle the
mixing in the neutralino-chargino sector to measure the supersymmetry
breaking masses for $SU(2)$ and $U(1)$ gauginos, $M_2$ for wino and
$M_1$ for bino, at future $e^+ e^-$ linear
colliders.\,\cite{Tsukamoto,e+e-SUSY} At the LHC, mass differences can
be measured well by identifying the end points in the decay
distributions.  Upon the assumption that the first and second
neutralinos are close to pure bino and wino eigenstates (which may be
cross-checked by other analyses of the data), the mass differences
also test the gaugino unification.\,\cite{Marjie}  Putting information
from both types of experiments, we will have three new numbers to deal
with.  This will provide us two more independent test if they unify at
the scale determined from the gauge coupling unification.

\begin{figure}[p]
  \begin{center}
    ~\hfill\psfig{file=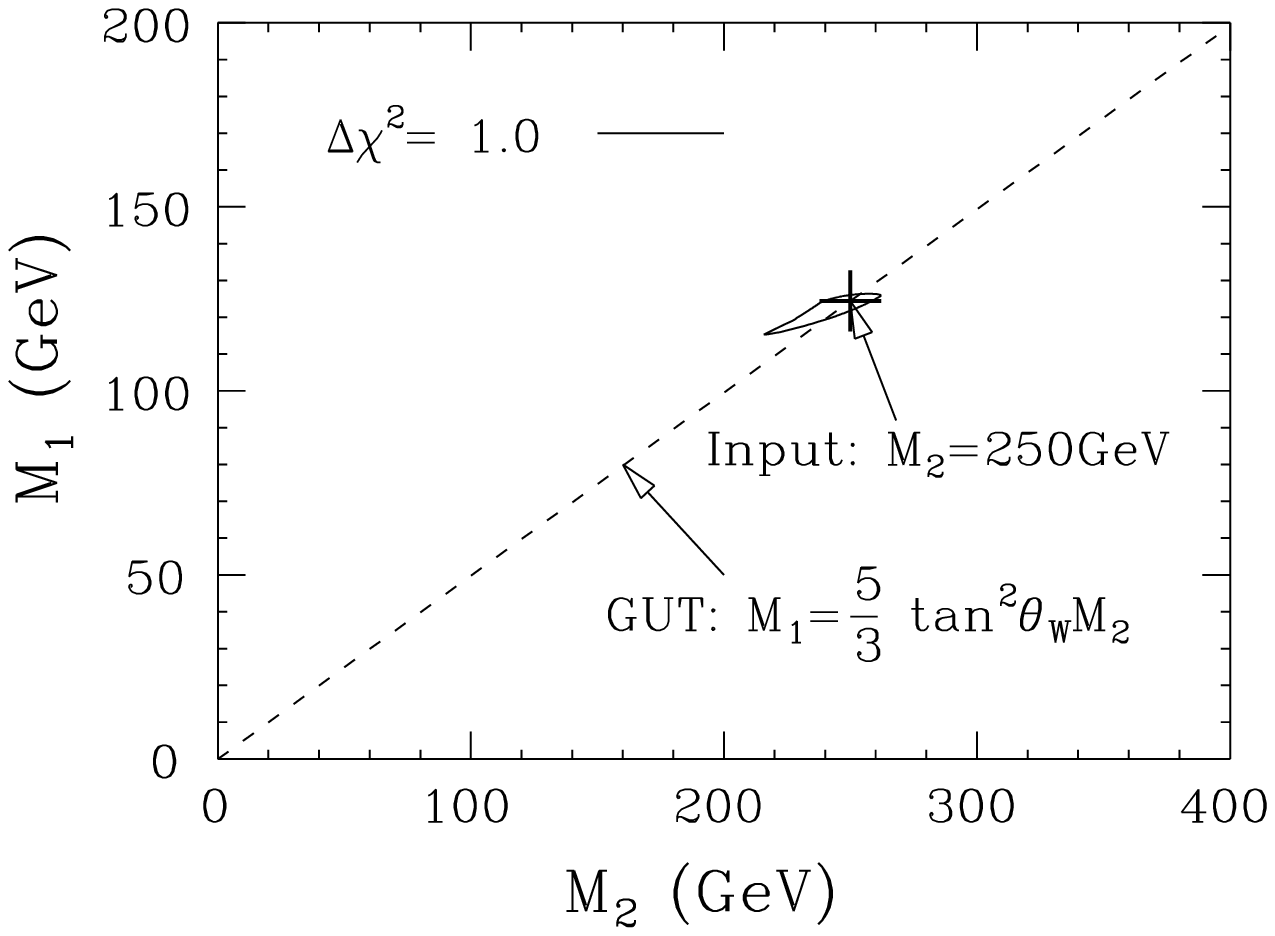,width=0.6\textwidth,angle=90}\hfill~\\
    ~\hfill\psfig{file=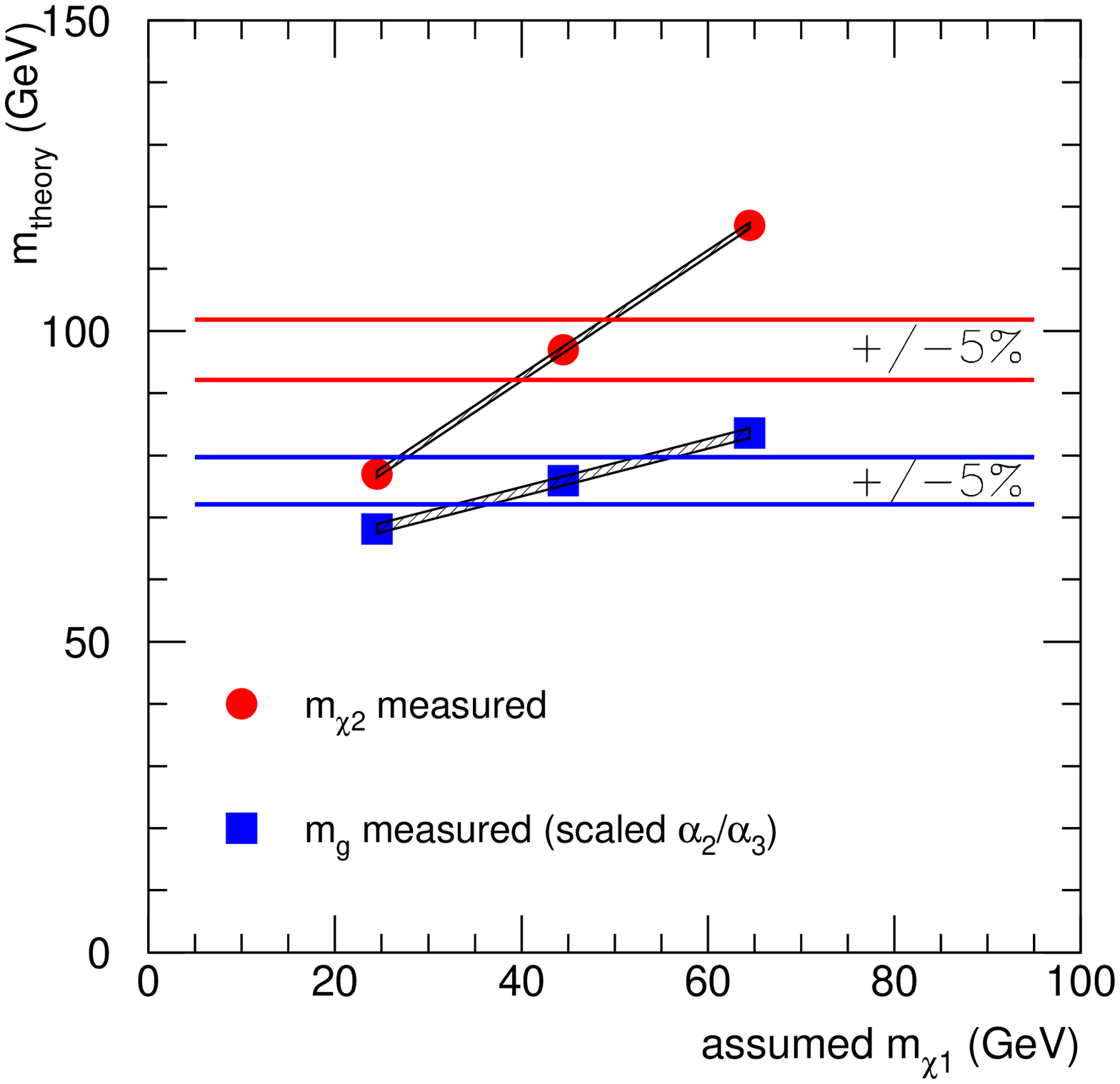,width=0.6\textwidth}\hfill~
  \end{center}
  \caption[gaugino]{Experimental tests of gaugino mass unification at
    a future $e^+ e^-$ collider\,\cite{Tsukamoto} and the LHC.\,\cite{Marjie}}
\end{figure}

It is an important question if the gaugino mass unification can be
spoiled even within GUT models.  One possible effect is the threshold
correction at the GUT scale.  Fortunately, there is no logarithmic
threshold corrections unlike to the gauge coupling constants\,\cite{HMG}
and hence the gaugino mass unification is a better prediction of GUT
than the gauge coupling unification.  The only way to spoil the
gaugino unification from the threshold corrections is to have
extremely large representations under the grand unified group with
highly non-universal trilinear and bilinear supersymmetry breaking
parameters.  Another possible effect is a higher dimension operator in
the gauge kinetic function which depends on the GUT-Higgs field, such
as $\int d^2 \theta \mbox{Tr}(\frac{\Sigma}{M} W_\alpha W^\alpha)$ we
discussed before, with an $F$-component VEV of the $\Sigma$
field.\,\cite{Tsukamoto} If we take $M$ at the reduced Planck scale and
$F_\Sigma \simeq m_{SUSY} M_{GUT}$, this operator generates an order
percent correction to the gaugino mass unification.  Note, however,
that the size of the $F$-component VEV tends to be only of
$m_{SUSY}^2$ in a wide class of supersymmetry breaking.\,\cite{KMY2,DPR}

On the other hand, there is a case where we may be fooled by the
apparent gaugino mass unification.  In the models of gauge mediated
supersymmetry breaking,\,\cite{DN,DNS,DNNS} the gaugino masses may
satisfy the same relation as the case with grand unification even
though the supersymmetry breaking gaugino masses have nothing to do
with the physics of grand unification.  However, this happens only
when the messenger fields fall into full $SU(5)$ multiplets and when
they acquire masses from the same field which has both $A$- and
$F$-component VEVs.  This is naturally expected in the GUT models,
even when the supersymmetry breaking is induced well below the
GUT-scale.  On the other hand, there is no reason for the messengers
to fall into full $SU(5)$ multiplets and acquire masses from the same
field if the theory is not grand unified.  Even though this remains as
a logically possibility that the data can fool us, the apparent
gaugino mass unification still strongly suggests grand unification.
One case which probably cannot be distinguished on the bases of
gaugino masses is the dilaton-dominant supersymmetry breaking in
superstring models.\,\cite{dilaton}  In this case, however, there is a
specific prediction on the ratio of the scalar masses (universal to
all scalars) and gaugino masses (universal to all gauginos) and can be
confronted to the data.

There are GUT-like models which do not lead to unified gaugino masses,
when the unified group is not simple.  One example is flipped
$SU(5)$.\,\cite{flipped} In this case, we expect to see the unification
of $M_2$ and $M_3$ at the scale where the gauge coupling constants
$\alpha_2$ and $\alpha_3$ meet, while $M_1$ may not.  This is an
interesting discriminator.  The other is the model of dynamical
GUT-breaking based on $SU(5)\times SU(3)\times
U(1)$.\,\cite{strongGUT,strongGUTgauginos} Here the gaugino masses do
not appear unified at all.

\subsection{Scalar Masses}

Under $SU(5)$ grand unified group, quarks and leptons belong to either
{\bf 10} or {\bf 5}$^*$ multiplets.  Under the same assumption that
the supersymmetry breaking masses respect grand-unified symmetry, we
can extrapolate the observed scalar masses to higher energies and see
if they unify at the same scale where the gauge couplings and gaugino
masses unify (if they do).  

The scalar mass unification will be an independent useful piece of
information beyond that from gauge couplings and gaugino masses.  One
probably very convincing case for grand-desert $SU(5)$ unification is
when both the gauge couplings and gaugino masses all unify at the same
scale, and also scalars in {\bf 10} and {\bf 5}$^*$ unify there but
with different masses.  On the other hand, the dilaton-dominated
supersymmetry breaking predicts the universal scalar mass, not
separate for ${\bf 10}$ and ${\bf 5}^*$, and a definite ratio of the
scalar mass to the gaugino mass ratio.

There is a possibility that we get fooled by an apparent unification
of gauge couplings and gaugino masses.  This happens, for instance,
when the supersymmetry breaking is induced by gauge mediation with
messengers in full $SU(5)$ multiplets which acquire both
supersymmetric and supersymmetry-breaking masses from a single field.
We argued in the previous section that such a case already strongly
suggests grand unification, but there remains a possibility that it is
not.  In this case, the scalar masses do not appear grand unified, and
provide a way to differentiate the conventional grand-desert $SU(5)$
GUT from gauge-mediated supersymmetry breaking.

Unlike the gaugino masses, the scalar mass spectrum is sensitive to
the pattern of GUT symmetry breaking.\,\cite{KMY1} Many different
patterns of scalar masses were discussed from GUT
models.\,\cite{Martin,Cheng-Hall} An important effect is that the
scalars which originally resided in the same GUT multiplet may acquire
different contributions from the $D$-term VEV when the rank of the
gauge group is reduced.\,\cite{D-term} For instance, all quarks and
lepton fields live in the same {\bf 16} multiplet under $SO(10)$; but
the breaking of $SO(10)$ to $SU(5)$ generally splits the {\bf 10} and
{\bf 5}$^*$ masses because of the $D$-term.  The $D$-term
contributions are determined solely by the gauge quantum numbers under
the broken gauge group and hence generation blind, and are safe from
the point of view of flavor-changing effects.  A complicated
superpotential interactions may modify the scalar masses as
well.\,\cite{KMY2,Tahoe,DPR} An extreme case is when the quarks and
leptons in the Standard Model come from different GUT multiplets; then
their scalar masses can be totally unrelated.\,\cite{Dimopoulos-Pomarol}
However, the constraint from the flavor-changing neutral currents and
smallness of Yukawa couplings for the first, second generation give us
a prejudice that the superpotential couplings, which can potentially
split the mass of first and second generation scalars in a
non-universal manner, are small.  Therefore, it is likely that the
scalar masses unify according to the patterns of the GUT symmetry
breaking, at least for the first and second generations.

\begin{figure}[t]
  \centerline{
  \psfig{file=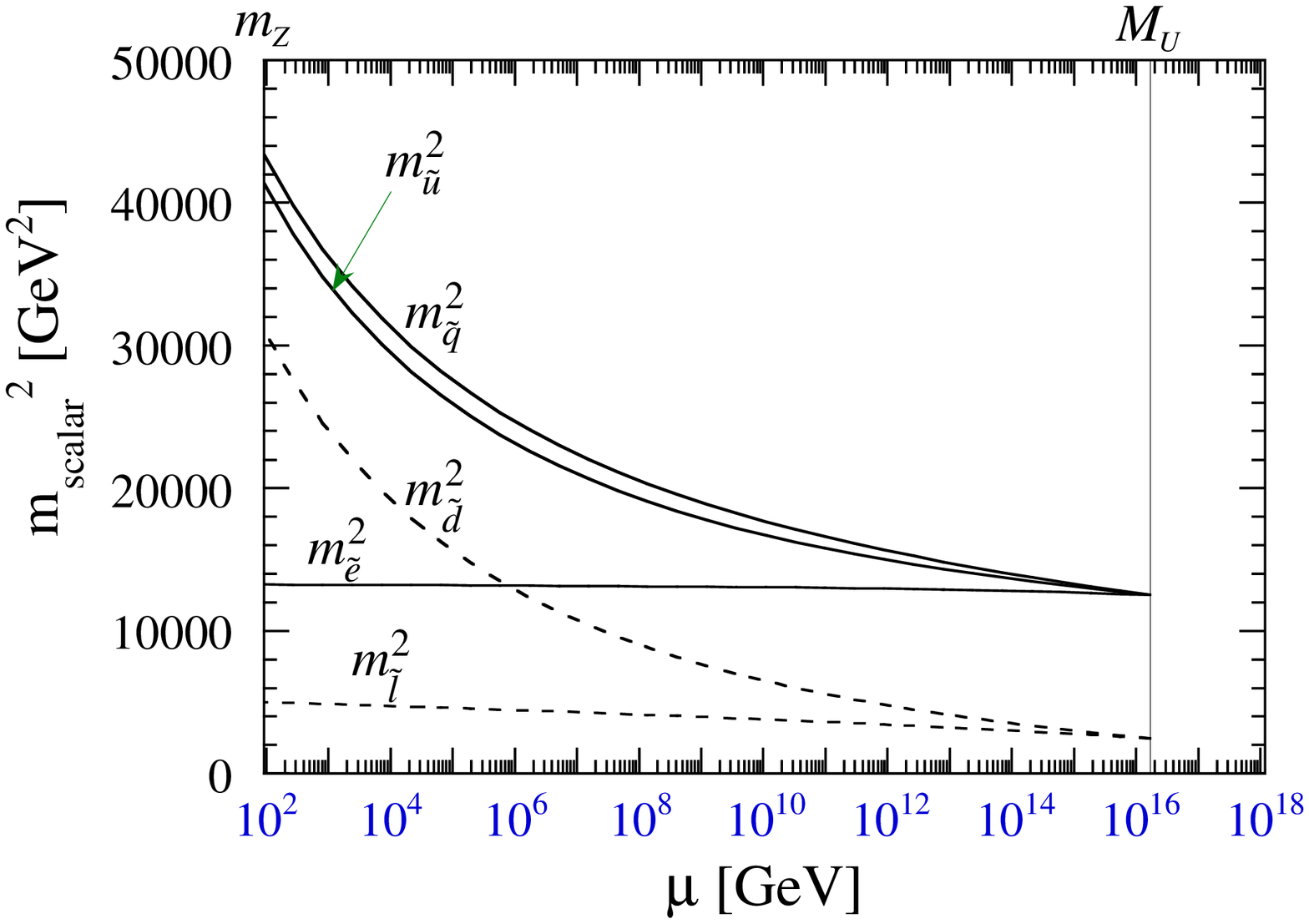,width=0.49\textwidth}
  \psfig{file=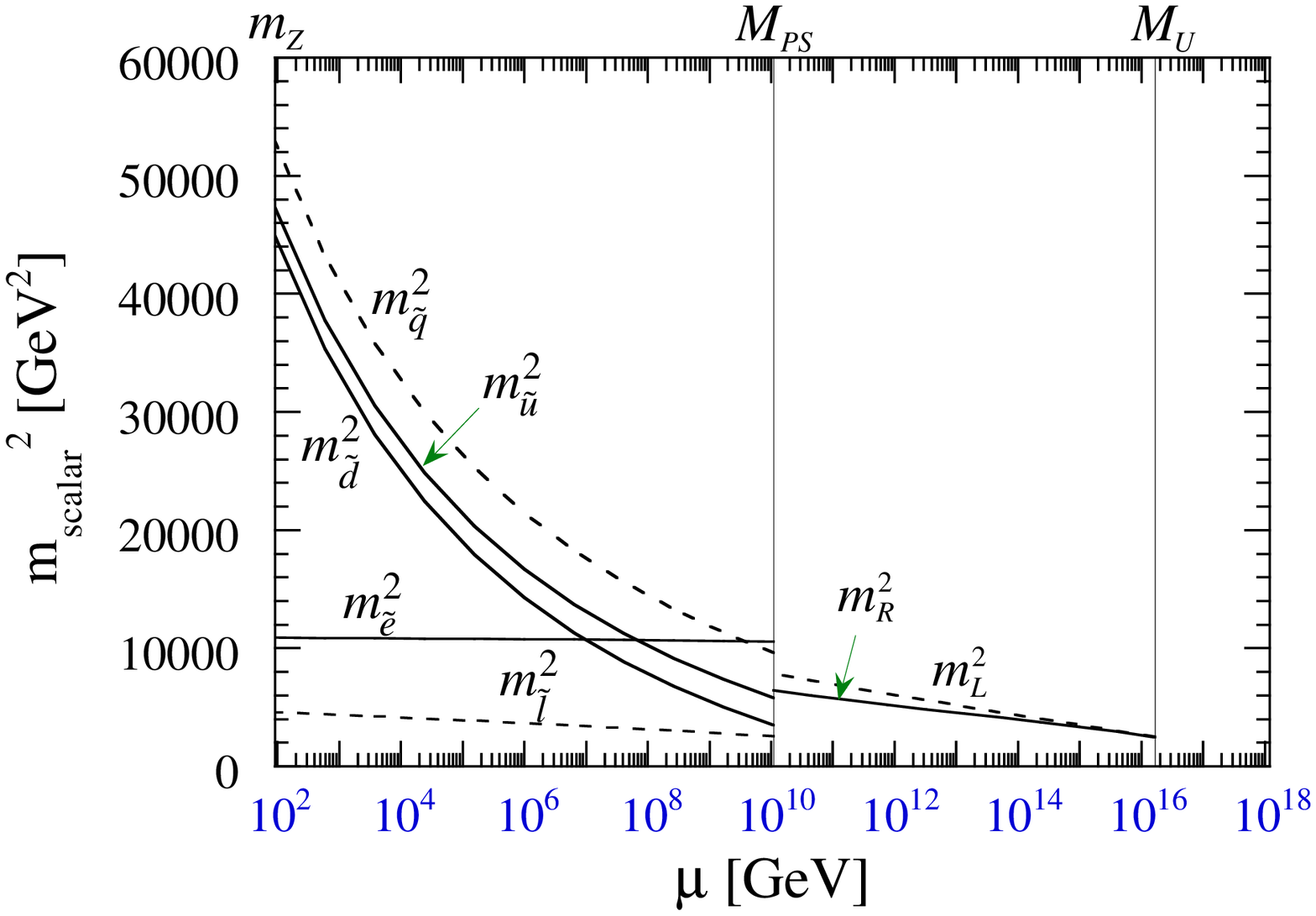,width=0.49\textwidth}
  }
  \caption[coupling]{The renormalization group evolution of the scalar
    masses in two models, $SU(5)$ GUT with grand desert and $SO(10)$
    GUT with an intermediate Pati--Salam symmetry used in
    Fig.~\ref{Fig:gauge}, which cannot be discriminated based on gauge
    coupling constants and gaugino masses.}
\label{Fig:scalar}
\end{figure}

By allowing $D$-term contributions but not $F$-term contributions
motivated by the above argument, one can try to fit the observed
scalar mass spectrum as a function of the symmetry breaking scales and
original supersymmetry breaking parameters.  For more complicated
symmetry breaking patterns, there are less relations and hence the
model is harder to test.  But still in many interesting symmetry
breaking patterns, there remain non-trivial relations among scalar
masses which can be confronted to data.\,\cite{KMY1,Martin,Cheng-Hall}
The Fig.~\ref{Fig:scalar} shows how the scalar masses acquire
different patterns at the electroweak scale between the grand-desert
$SU(5)$ and the toy Pati--Salam model, which could not be
distinguished based on the gauge coupling constants and the gaugino
masses.  Therefore the role of gaugino mass unification and scalar
mass unification are complimentary; the former gives a
model-independent test of the grand unification, while the latter
selects out particular symmetry breaking patterns and their energy
scales.\,\cite{KMY1}

Even in the case of non-simple GUT groups, the scalar masses still
give us useful information.  In flipped $SU(5)$, different sets of
fields $(Q, d^c, \nu)$ belong to {\bf 10} and $(L, u^c)$ to {\bf
  5}$^*$, and $e^c$ is a singlet by itself.  Therefore, there is still
the scalar-mass unification of $Q$ and $d^c$, and $L$ and $u^c$
separately.  In the model of dynamical GUT-breaking based on
$SU(5)\times SU(3)\times U(1)$,\,\cite{strongGUT}, still all matter
fields belong to the ordinary ${\bf 10}+{\bf 5}^*$ multiplets and the
pattern of scalar masses is the same as in the grand-desert $SU(5)$.

Experimentally, measurement of scalar masses is also feasible.  At an
$e^+ e^-$ collider, a well-defined kinematics allows a simple
kinematic fit to the decay distributions to extract the mass of the
parent scalar particle.  This comment applies both to the
sleptons\,\cite{Tsukamoto,e+e-SUSY} and squarks\,\cite{Feng-Finnell} using
beam polarizations, as long as they are within the kinematic reach.
At the LHC, the mass differences can be measured well as before;
especially when the second neutralino decays into on-shell sleptons,
one has a high rate and the mass difference between the slepton and
the lightest neutralino is measured very well.  Many other mass
patterns also allow certain mass differences to be measured accurately
at a few percent level.\,\cite{Marjie} It is quite imaginable that the
spectroscopy of superparticles will be the main experimental project
of the next decade.

\section{Proton Decay}

Proton decay has been virtually the only direct probe of the physics
at the GUT-scale and discussed extensively in the literature.  The
original idea is that the gauge bosons in $SU(5)$ GUT cause
transitions between quarks and leptons in the same $SU(5)$ multiplets
and hence allow proton to decay.  Assuming that the quarks and leptons
of the first generation belong to the same $SU(5)$ multiplets, the
exchange of the heavy $SU(5)$ gauge boson generates an operator
\begin{equation}
  {\cal L} = \frac{1}{M_V^2} u u d e,
\end{equation}
which gives rise to a decay $p \rightarrow e^+ \pi^0$.  The current
experimental bound excludes the process for heavy gauge bosons
approximately up to $1.5\times 10^{15}$~GeV,\,\cite{Warsaw} where we
estimated the bound conservatively.\,\cite{HMY} Because the operator has
a suppression by two powers of a high mass scale, the proton decay
rate is suppressed by the fourth power in the mass scale $\Gamma_p
\propto m_p^5/M^4$.  It is not easy to extend the experimental reach
on $M$.  SuperKamiokande will probably extend the limit on the
lifetime by a factor of 30 beyond the current one, which translates to
a modest improvement by a factor of 2.3 on the GUT-scale.  ICARUS may
reach the mass scale of the supersymmetric GUT or $10^{16}$~GeV.

\subsection{$D=5$ operators}

The important and novel feature in supersymmetric models is that there
are operators of dimension-five which violate baryon and lepton
numbers and hence can cause proton decay.\,\cite{SYW}\footnote{In this
  discussion, we assume that there is no dimension-{\it four}\/
  operators which violate baryon and lepton numbers.  Such operators
  are conveniently forbidden by imposing the $R$-parity.}  For
instance, the following operator is possible in the superpotential:
\begin{equation}
  W = \frac{\lambda}{M} (Q_1 Q_1) (Q_2 L_i),
\end{equation}
where the subscript refers to the generation, and $\lambda$ is a
coupling constant.  The operator involves squarks and sleptons, which
need to be converted to quarks or leptons by a loop diagram.  The
proton decay rate therefore scales as $\Gamma_p \propto m_p^5
\lambda^2/(16\pi^2)^2 M^2 m_{SUSY}^2$ where $m_{SUSY}$ is the mass
scale of superparticles.  As a result, the reach in the energy scale
is drastically improved.  The current experimental limit does not
allow $M$ below $10^{24}$~GeV if $\lambda \sim 1$.  Therefore, we are
sensitive to even Planck-scale suppressed operators which, actually,
are excluded with $O(1)$ couplings.

It is interesting that the dimension-five operators necessarily 
involve quark superfields of different generations (at least two).  
This is a simple consequence of the Bose symmetry among superfields 
and the Standard Model gauge invariance.  The interesting consequence 
of this fact is that the proton (or neutron) decay modes 
preferentially involve kaons in the final state, such as $p 
\rightarrow K^{+} \bar{\nu}_{\mu}$ as predicted to be dominant in the 
minimal $SU(5)$ GUT.  If the dominant proton decay mode will be seen 
to involve kaons, it is likely to be a consequence of dimension-five 
operators possible only in supersymmetric theories.

\subsection{Minimal SUSY $SU(5)$}

In the minimal SUSY $SU(5)$ GUT,\,\cite{DGS} the dimension-five
operators are generated by the exchange of color-triplet Higgs
supermultiplet.  The Yukawa couplings of quarks to the Higgs doublets
are known from the quark masses, and the couplings to the
color-triplet Higgs ($SU(5)$ partner of the doublets) are the those to
the Higgs doublets at the GUT-scale because of the $SU(5)$ invariance.
Therefore, there is little freedom in this model and the size of the
dimension-five operators is completely fixed except possible relative
phases which become unobservable below the GUT-scale.\,\cite{ACP} The
mass of the color-triplet Higgs at first appears to be a free
parameter.  However, the gauge unification constrains its mass through
its threshold correction.\,\cite{HMY} At the one-loop level of the
renormalization group equations, one obtains
\begin{eqnarray}
(3 \alpha_2^{-1} - 2 \alpha_3^{-1} - \alpha_1^{-1}) (m_Z)
        &=& \frac{1}{2\pi} \left\{ 
                \frac{12}{5} \, \ln \frac{M_H}{m_Z}
                - 2 \, \frac{m_{SUSY}}{m_Z} \right\},
                        \label{MHC}
                        \rule[-0.77cm]{0cm}{1.7cm} \label{eq:HMY1}
                        \\
(5 \alpha_1^{-1} - 3 \alpha_2^{-1} - 2 \alpha_3^{-1}) (m_Z)
        &=& \frac{1}{2\pi} \left\{
                12 \, \ln \frac{M_V^2 M_\Sigma}{m_Z^3}
                + 8 \ln \frac{m_{SUSY}}{m_Z} \right\}.
                \label{MGUT}
\end{eqnarray}
Here, $m_{SUSY}$ stands for some weighted average of the superparticle
masses.  Once the mass spectrum of the superparticle is measured, one
can determine $m_{SUSY}$ in the above formulae, and then extract the
mass of the colored Higgs $M_{H_C}$, and a combination of $M_V$ and
$M_\Sigma$ from the renormalization group equations.  In fact, the
measured $\alpha_{s}$ is smaller than the preferred value from the GUT
and as a result prefers a low-value of color-triplet Higgs mass.
Since various $\alpha_{s}$ measurements basically converged recently
to $\alpha_{s} (m_{Z}) = 0.118 \pm 0.003$, the minimal $SU(5)$ model
is almost excluded\,\cite{minimal-excluded,Warsaw} unless extreme parameters
are chosen for gauginos (preferentially light) and squarks
(preferentially heavy).  Given the uncertainties in the superparticle
spectrum, it is hard to announce the definite exclusion of the model.
Once the superparticles are found, however, we will be able to make a
final word on the model, assuming the current value of $\alpha_{s}$
persists and the SuperKamiokande will not find proton decay.

\begin{figure}[t]
  \begin{center}
    \leavevmode
    \psfig{file=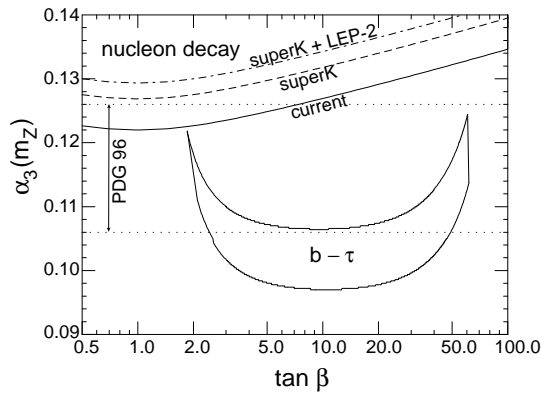,width=0.6\textwidth}
  \end{center}
  \caption[pdecay]{Excluded region on $(\tan\beta, \alpha_s (m_Z))$
    plane from nucleon decay based on very conservative assumptions
    using constraints on superparticle spectrum from LEP-1
    only.\,\cite{Warsaw} Expected improvements from SuperKamiokande and
    LEP-2 are also shown.  The range shown for $\alpha_s (m_Z)$ from
    PDG96 is two-sigma range.  The preferred region from $b-\tau$
    unification is also shown for $m_t=176$~GeV as a crescent-shaped
    region.}
\end{figure}

\subsection{Non-minimal SUSY-GUT}

There are many good reasons to discuss extensions of the minimal SUSY
SU(5) GUT.  Among them, there are two points directly relevant to the
nucleon decay.  (1) The triplet-doublet splitting problem.  In minimal
SUSY SU(5) GUT, one needs to fine-tune independent parameters at the
level of $10^{-14}$ to keep Higgs doublets light while making the
color-triplet Higgs heavy.  (2) The wrong fermion mass relations.  It
predicts $m_s = m_\mu$ and $m_d = m_e$ at the GUT-scale, which are off
from the phenomenologically preferred Georgi--Jarlskog relations $m_s =
m_\mu/3$, $m_d = 3 m_e$.  

Solutions to the above-mentioned problems modify the predicted rate and
branching ratios of the nucleon decay.  One possible attempt to obtain
Georgi--Jarlskog relations is to use the SU(5)-adjoint Higgs to
construct an effective {\bf 45} Higgs doublets as composites of ordinary
Higgs doublets in {\bf 5} and the adjoint.  This modification leads to a
factor-of-two enhancement in the amplitude; a factor of four in the
rate.\,\cite{INS}  The relative branching ratios can be also different.
It remains true that the $K^{+,0} \bar{\nu}_\mu$ modes are the dominant
ones, while the $K^0 \mu^+$ mode may be much less suppressed than in the
minimal SU(5).\,\cite{Antaramian,BB}  the proton decay in $SO(10)$
models with realistic fermion mass texture has been also discussed
extensively.\,\cite{Rabyproton,BPW} 

There are various proposals to solve the triplet-doublet splitting
problem, which lead to different nucleon decay phenomenology.  I
discuss three of them here.  (1) The missing partner model,\,\cite{MNTY}
(2) Dimopoulos--Wilczek--Srednicki mechanism,\,\cite{DWS} and (3)
flipped SU(5) model.\,\cite{flipped}

In the missing partner model, one employs {\bf 75} representation to
break SU(5) instead of the adjoint {\bf 24}, and further introduces
{\bf 50} and {\bf 50}$^*$ representations which mix with the
color-triplet Higgs to make them massive.  Since the model involves
such large representations, the size of the GUT-scale threshold
corrections are significantly larger than that in the minimal model.
And the correction changes the determination of the color-triplet
Higgs mass as done in Eq.~(\ref{eq:HMY1}), and the measured values of
the gauge coupling constants prefer larger $M_{H_C}$ than in the
minimal model.\,\cite{Yamada} In this case the proton decay rates are
much more suppressed, by a few orders of magnitudes.  One drawback of
the model is that it becomes non-perturbative well below the Planck
scale due to large representations and one needs to complicate the
model further to keep it perturbative.\,\cite{missingPQ} It is worth to
recall that the minimal SU(5) model is marginally allowed only with
very conservative assumptions.  Even though there is an additional
suppression to the proton decay rate in this class of models, the
decay rate may still well be within the reach of superKamiokande
experiment.

The mechanism proposed by Dimopoulos, Wilczek and further by Srednicki
employs SO(10) unification with Higgs fields in adjoint and symmetric
tensor representations which naturally keep Higgs doublets light.
However, their model breaks SO(10) only to SU(3)$\times$SU(2)$_L
\times$SU(2)$_R$ and has to be extended to achieve the desired symmetry
breaking down to the standard model gauge group.  One of such extensions
by Babu and Barr\,\cite{BB0} eliminates $D=5$ entirely; but it involves
rather complicated Higgs sector, and one needs to forbid some allowed
interactions in the superpotential arbitrarily.  A later attempt to
guarantee the special form of the superpotential by symmetries did not
eliminate the $D=5$ operators entirely, but resulted in a weak
suppression of the operators.  Again in view of the very marginal
situation in the minimal model, the decay rate could be within the reach
of the superKamiokande.

The flipped SU(5) model solves the triplet-doublet splitting problem in
a way that it also eliminates the $D=5$ operators entirely.  A possible
problem with this model is that the gauge unification becomes more or
less an accident rather than a prediction.  On the other hand, the
elimination of the $D=5$ operator is a natural consequence of the
structure of the Higgs sector, and is rather a robust prediction of the
model except the Planck-scale effects which will be discussed below.  An
interesting feature of the model is that the GUT-scale is determined by
$\alpha_2$ and $\alpha_3$ and hence can be {\it lower}\/ than the scale
in the minimal SU(5) which is determined by $\alpha_2$ and $\alpha_1$.
Since the model does not predict the relation between $\alpha_1$ and
$\alpha_{SU(5)}$, $\alpha_1$ does not need to meet with the other
coupling constants at the same scale.  Therefore, the GUT-scale can be
as low as $M_{GUT}^{\rm flipped} = 4$--$20\times 10^{15}$~GeV.  If the
$M_{GUT}$ is at the low side within this range, the $D=6$ operator may
be observable in the $\pi^0 e^+$ mode,\,\cite{ELN} since the superKamiokande
is expected to extend the reach by a factor of 30.  

Certain models of direct gauge mediation\,\cite{direct} also have
$SU(5)$ group broken below the typical GUT-scale and can lead to
dimension-six proton decay at a rate observable at superKamiokande.
There is also a variant of missing-partner model with dimension-six
proton decay within the reach of superKamiokande.\,\cite{HNY}

\subsection{Planck-scale Operators}

Planck-scale physics may generate $D=5$ operators suppressed by the
reduced Planck scale $M_* = 2\times 10^{18}$~GeV.  Even when there is
no color-triplet Higgs, such as in string compactifications which
breaks the gauge group down to the standard model (with possible U(1)
factors) directly, the higher string excitations may give rise to
effective non-renormalizable $D=5$ operators which break baryon-
and/or lepton-number symmetries.  For $D=5$ operators which involve
first- and second-generation fields, $1/M_*$ suppression is far from
enough: one needs a coupling constant of order $10^{-7}$ to keep the
nucleons stable enough as required by experiments.

It is a serious question in supersymmetry phenomenology why the
Planck-scale $D=5$ operators are so much suppressed.  One possibility
is to forbid them by employing discrete gauge
symmetries\,\cite{discreteB} which are believed to be respected by
quantum gravitational effects unlike global symmetries.  In this case,
there is no baryon-number-violating $D=5$ operator from Planck-scale
physics and we do not have any handle on it.  A different type of
solution is probably more interesting: the $D=5$ operators are
suppressed because of the same reason why the Yukawa couplings of
light generations are suppressed.\,\cite{MK} One way to understand why
the Yukawa couplings are so small, such as $10^{-6}$ for the case of
the electron, may be a natural consequence of an approximate flavor
symmetry.  If a flavor symmetry exists and is only weakly broken to
explain smallness of the Yukawa couplings, the same flavor symmetry
can well suppress the $D=5$ operators at the Planck-scale.  The $D=5$
operators with such a flavor origin may have very different flavor
structure from those in the GUT models, and may lead to quite
different decay modes like $p\rightarrow K^0 e^+$.

Suppression of Planck-scale $D=5$ operators based on certain flavor
symmetries were discussed.\,\cite{Nirproton,S3}  For instance the
$S_3^3$ model\,\cite{S3} explains the
hierarchical Yukawa matrices as a consequence of
sequential breaking of the flavor symmetry while the symmetry
preserves sufficient degeneracy among the squarks and sleptons to
suppress flavor-changing neutral currents.  It happens that the
flavor symmetry in this model also suppresses $D=5$ operators to the
level of about 1/9 of the minimal $SU(5)$ model, so that it can well be
within the reach of superKamiokande.\,\cite{S3q}  What is particularly
interesting in this model is that it predicts $p \rightarrow K^0 e^+$
as the {\it dominant}\/ mode over the $K^+ \bar{\nu}$, while
$n\rightarrow K^0 \bar{\nu}_e$ is the dominant mode in neutron decay
with a comparable rate.  In general, decay modes of proton, if
observed, will provide interesting information on the flavor
physics.\,\cite{BB,Antaramian,BPW} 

\section{Flavor Physics}

Another interesting topic is how well we will be able to understand
the origin of flavor, fermion masses and mixing based on the study of
supersymmetry possible at the electroweak scale.  Unlike the case of
grand unification and proton decay, the answer to this question
depends heavily on what the true story is.  

\subsection{Neutrino Physics}

An analogue of proton decay discussed in the previous section is a
consequence of flavor physics suppressed by powers of the mass scale,
such as the neutrino mass via the seesaw mechanism.\,\cite{seesaw}  The
neutrino masses are generated from their Dirac masses $m_D$ with
right-handed neutrinos neutral under the Standard Model gauge groups
and their Majorana masses $M$ which violate the lepton numbers by two
units.  The one-generation case is given by a two-by-two mass matrix
\begin{equation}
  {\cal L}_{\it mass} = \frac{1}{2} (\nu, N^c)
  \left( \begin{array}{cc} 0 & m_D \\ m_D & M \end{array} \right)
  \left( \begin{array}{cc} \nu \\ N^c \end{array} \right),
\end{equation}
where $\nu$ is the Weyl field of the left-handed neutrino, and $N$ the
right-handed neutrino.  The Lagrangian is written in terms of the
charge-conjugated Weyl spinor $N^c$ with the same chirality as the
left-handed field $\nu$.  After diagonalization of the mass matrix,
one obtains a mass for the left-handed neutrino of $m_D^2/M$, which is
power suppressed.  This mechanism naturally explains why the neutrino
masses are so small, if finite, and leaves imprint of short-distance
physics in the pattern of neutrino masses and mixings.  

Let us emphasize that this is the area where a dramatic progress is
likely to be made in the next few years, from SuperKamiokande (together
with neutrino beam from KEK), CHORUS, NOMAD, KARMEN, SNO, BOREXINO,
MINOS, and more.  Even though supersymmetry does not necessarily help
to study the physics at the scale of right-handed neutrinos, many
flavor models in supersymmetry predict interesting patterns on
neutrino masses.\,\cite{neutrino}  We will certainly be making
selections on different flavor models based on neutrino physics if
finite neutrino masses and their mixings will be established.

\subsection{Flavor-Changing Neutral Currents}

As discussed in other chapters, there are severe constraints on the
superparticle masses and mixings from the flavor-changing neutral
currents (FCNC).  There are broadly three categories of models which
naturally suppress the FCNC.  (1) Flavor symmetry enforces the
squarks, sleptons to be degenerate,\,\cite{degenerate} or aligns their
mass basis to that of the down-quark, charged-lepton mass
basis.\,\cite{align} (2) The string theory generates universal scalar
mass.\,\cite{dilaton} (3) The supersymmetry breaking is generated in a
flavor-blind fashion below the scale of flavor
physics.\,\cite{DN,DNS,DNNS}

In cases (1) and (2), there may be interesting imprints of flavor
physics in the small mixing between squarks and sleptons.  The case of
GUT also belongs to the category: the large top Yukawa coupling above the
GUT-scale may affect the slepton masses-squared with small mixing
between, for instance, selectron and smuon.\,\cite{HKR,BH1}  The search for rare
decays such as $\mu\rightarrow e\gamma$,\,\cite{BH1,muegamma}
$K$-physics,\,\cite{BHforK} or electric dipole moments of electron or
neutron\,\cite{EDM} may reveal the imprints of flavor physics in
scalar masses.  At present, the main uncertainty in the quantitative
analysis is the mass of superparticles.  Once they are measured,
however, we can try to extract the mixing effects in the scalar mass
matrices from the FCNC data.

\begin{figure}[t]
  \begin{center}
    \leavevmode
    \psfig{file=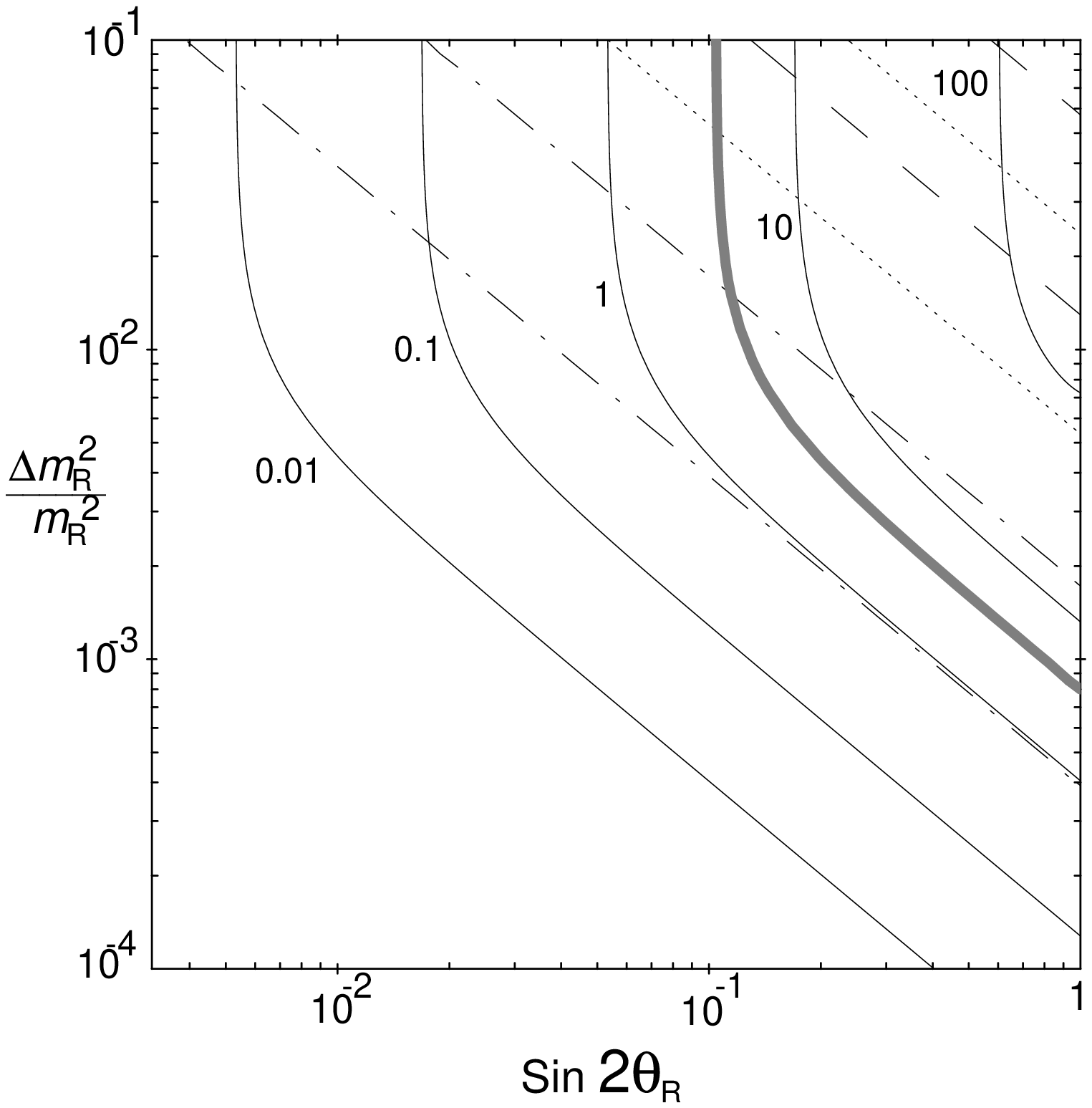,width=0.49\textwidth}
    \psfig{file=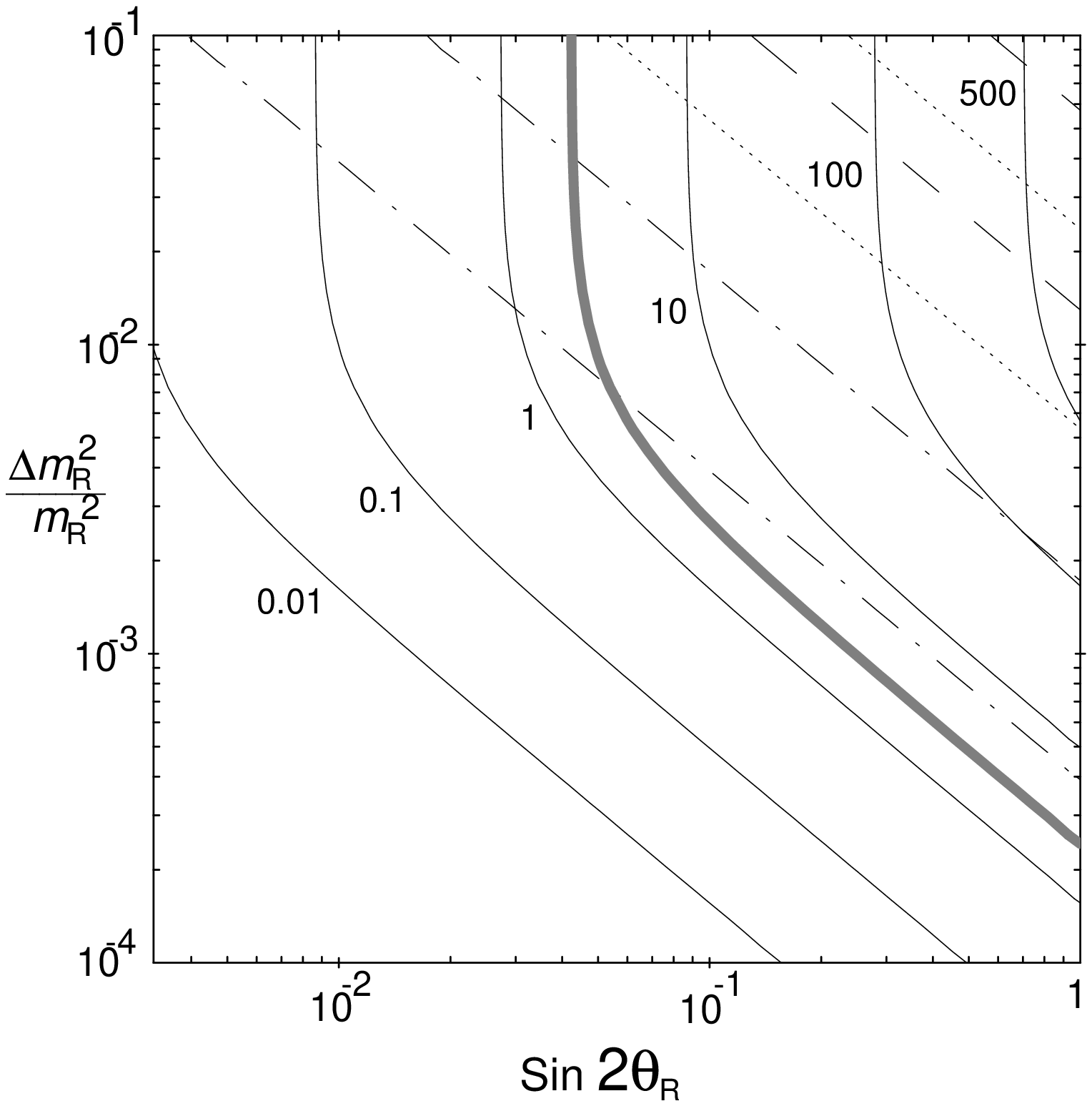,width=0.49\textwidth}
  \end{center}
  \caption{Contours of constant $\sigma (e^+e^-_R\to e^{\pm} \mu^{\mp} 
    \tilde{\chi}^0\tilde{\chi}^0 )$ (solid) and $\sigma (e^-_R e^-_R
    \to e^- \mu^- \tilde{\chi}^0\tilde{\chi}^0 )$ (solid) in fb for
    $e^+ e^-$ or $e^- e^-$ linear colliders, with $\protect\sqrt{s} =
    500~\mbox{GeV}$, $m_{\tilde{e}_R}, m_{\tilde{\mu}_R} \approx
    200~\mbox{GeV}$, and $M_1 = 100~\mbox{GeV}$ (solid).  The thick
    gray contour represents the experimental reach in one year.
    Constant contours of $B(\mu \to e\gamma)=4.9\times 10^{-11}$ and
    $2.5\times 10^{-12}$ are also plotted for degenerate left-handed
    sleptons with mass 120 GeV and $\tilde{t} \equiv -(A + \mu
    \tan\beta)/\bar{m}_R = 0$ (dotted), 2 (dashed), and 50
    (dot-dashed), with left-handed sleptons degenerate at 350
    GeV.}
  \label{Fig:LFV}
\end{figure}

An interesting case where the flavor-mixing effects in scalar masses
can be probed at colliders was discussed.\,\cite{ACFH}  Analogous to
neutrino oscillation, a selectron produced from an $e^+ e^-$ or $e^-
e^-$ collider can oscillate to a smuon as a result of the flavor
mixing and is detected as $e\mu$ final state (see Fig.~\ref{Fig:LFV}).  

In the case (3) where the scalar masses are generated in a
flavor-blind fashion below the flavor physics scale, such as in the
models of low-energy gauge mediation, we unfortunately may not learn
about the origin of flavor from the study of flavor signatures at the
electroweak scale.

\section{Conclusion}

Experiments at the electroweak scale will remove the cloud which masks
the physics at yet shorter distance scales.  If supersymmetry turns
out to be the mechanism of stabilizing the electroweak scale, we will
have a wealth of new data on superparticle spectroscopy.  Combined with data
on proton decay, neutrino physics, and FCNC, we will obtain
useful information on physics such as grand unification, string,
flavor physics.  At this point it is just a dream; but we may be able
to glimpse the physics at the shortest possible distance scales by
this program, which is nothing but the goal of particle physics after all.

\section*{Acknowledgements} This work was supported in part by the U.S. 
Department of Energy under Contracts DE-AC03-76SF00098, in part by the 
National Science Foundation under grant PHY-95-14797, and also 
by Alfred P. Sloan Foundation.


\begin{thebibliography}{99}
  \bibitem{ICEPP} H. Murayama, Invited Talk at the ICEPP
    Symposium ``From LEP to the Planck World,'' University of Tokyo,
    Dec 17--18, 1992.  In Proceedings of the ICEPP Symposium ``From
    LEP to the Planck World,'' eds. K.~Kawagoe and T.~Kobayashi,
    UT-ICEPP 93-12, TU-451, 11pp.  
    
  \bibitem{FPMT} J.L. Feng, M.E. Peskin, H. Murayama, and X. Tata,
    \Journal{\PRD}{52}{1418}{1995}.
    
  \bibitem{FNT} M. M. Nojiri, K. Fujii, and T.Tsukamoto,
    \Journal{\PRD}{54}{6756}{1996}.

  \bibitem{HMY} J. Hisano, H. Murayama, and T. Yanagida,
    {\it Phys. Rev. Lett.} {\bf 69}, 1014, (1992); {\it Nucl. Phys.}\/
    {\bf B402}, 46 (1993). 
    
  \bibitem{Hall-Sarid} L. J. Hall and U. Sarid,
    \Journal{\PRL}{70}{2673}{1993}.

  \bibitem{Nath} T. Dasgupta, P. Mamales, and P. Nath,
    \Journal{\PRD}{52}{5366}{1995}. 

  \bibitem{DKR} N.G. Deshpande, E. Keith, and T.G. Rizzo,
    \Journal{\PRL}{70}{3189}{1993}. 

  \bibitem{KMY1}
    Y. Kawamura, H. Murayama, and M. Yamaguchi, {\it Phys. Lett.}\/
    {\bf B324}, 52 (1994). 

  \bibitem{Tsukamoto} T. Tsukamoto, K. Fujii, H. Murayama,
    M. Yamaguchi, and Y. Okada, \Journal{\PRD}{51}{3153}{1995}.
    
  \bibitem{e+e-SUSY} J. L. Feng and M. J. Strassler,
    \Journal{\PRD}{51}{4661}{1995}; \Journal{\PRD}{55}{1326}{1997}; H.
    Baer, R. Munroe, and X. Tata, \Journal{\PRD}{54}{6735}{1996};
    \Journal{Erratum-ibid}{56}{4424}{1997}.

  \bibitem{Marjie} I. Hinchliffe, F.E. Paige, M.D. Shapiro, J.
    Soderqvist, and W. Yao, \Journal{\PRD}{55}{5520}{1997}.

  \bibitem{HMG} J. Hisano, H. Murayama, and T.
    Goto, {\it Phys. Rev.}\/ {\bf D49}, 1446 (1994).

  \bibitem{KMY2} Y. Kawamura, H. Murayama, and M.
    Yamaguchi, {\it Phys. Rev.}\/ {\bf D51}, 1337 (1995).

  \bibitem{DPR} A. Pomarol and S. Dimopoulos,
    \Journal{\NPB}{453}{83}{1995}; R. Rattazzi,
    \Journal{\PLB}{375}{181}{1996}. 

  \bibitem{DN} M. Dine and A.E. Nelson, {\it Phys. Rev.}\/ {\bf D48}, 1277
    (1993).

  \bibitem{DNS} M. Dine, A.E. Nelson and Y. Shirman, {\it Phys. Rev.}\/
    {\bf D51}, 1362 (1995).

  \bibitem{DNNS} M. Dine, A.E. Nelson, Y. Nir and Y. Shirman, {\it Phys.
      Rev.}\/ {\bf D53}, 2658 (1996).

  \bibitem{flipped} S. M. Barr, \Journal{\PLB}{112}{219}{1982}; J.P.
    Derendinger, J. E. Kim, and D.V. Nanopoulos,
    \Journal{\PLB}{139}{170}{1984}; I. Antoniadis, J. Ellis, J.S.
    Hagelin, and D.V. Nanopoulos, \Journal{\PLB}{194}{231}{1987}.

  \bibitem{strongGUT} T. Yanagida, \Journal{\PLB}{344}{211}{1995}; 
    
  \bibitem{strongGUTgauginos} N. Arkani-Hamed, H.-C. Cheng, and T.
    Moroi, \Journal{\PLB}{387}{529}{1996}.

  \bibitem{dilaton} V. S. Kaplunovsky and J. Louis,
    \Journal{\PLB}{306}{269}{1993}; A. Brignole, L.E. Ibanez, and
    C. Munoz, \Journal{\NPB}{422}{125}{1994},
    \Journal{Erratum-ibid.}{436}{747}{1995}; A. Brignole, L.E. Ibanez,
    C. Munoz, and C. Scheich, \Journal{\ZPC}{74}{157}{1997}.

  \bibitem{Martin} C. Kolda, and S. P. Martin,
    \Journal{\PRD}{53}{3871}{1996}. 

  \bibitem{Cheng-Hall} H.C. Cheng, and L.J. Hall,
    \Journal{\PRD}{51}{5289}{1995}. 
  
  \bibitem{D-term} M.~Drees, {\it Phys. Lett.}\/ {\bf 181B}, 279 (1986);
    J.S.~Hagelin and S.~Kelley, {\it Nucl. Phys.}\/ {\bf
      B342}, 95 (1990);
    A.E.~Faraggi, J.S.~Hagelin, S.~Kelley, and D.V.~Nanopoulos, {\it
      Phys.  Rev.}\/ {\bf D45} 3272 (1992).
  
  \bibitem{Tahoe} H. Murayama, Invited plenary talk given at 4th
    International Conference on ``Physics Beyond the Standard Model,''
    Lake Tahoe, CA, 13-18 Dec 1994.  Proceedings, eds. by J. Gunion,
    T. Han, J. Ohnemus, World Scientific, 1995.


  \bibitem{Dimopoulos-Pomarol} S. Dimopoulos and A. Pomarol,
    \Journal{\PLB}{353}{222}{1995}. 
  
  \bibitem{Feng-Finnell} J. L. Feng and D. E. Finnell,
    \Journal{\PRD}{49}{2369}{1994}. 
  
  \bibitem{minimal-excluded} J. Hisano, T. Moroi, K. Tobe, and
    T. Yanagida, \Journal{\MPLA}{10}{2267}{1995}; J. Bagger,
    K. Matchev, and D. Pierce, \Journal{\PLB}{348}{443}{1995}.

  \bibitem{Warsaw} H. Murayama, Invited talk presented at 28th
    International Conference on High-energy Physics (ICHEP 96),
    Warsaw, Poland, 25-31 Jul 1996.  Published in the Proceedings of
    the 28th International Conference on High Energy Physics, {\it
      eds.}\/, Z.~Ajduk and A. K. Wroblewski, World Scientific, 1997,
    pp. 1377--1382.

  \bibitem{SYW} N. Sakai and Tsutomu Yanagida, 
    \Journal{\NPB}{197}{533}{1982}; S. Weinberg, 
    \Journal{\PRD}{26}{287}{1982}.

    
  \bibitem{DGS} S. Dimopoulos and H. Georgi,
    \Journal{\NPB}{193}{150}{1981}; N. Sakai,
    \Journal{\ZPC}{11}{153}{1981}.

  \bibitem{ACP} P.~Nath and R.~Arnowitt, \Journal{\PRD}{38}{1479}{1988}.

  \bibitem{direct} H. Murayama, {\it Phys. Rev. Lett.}\/ {\bf 79},
    18 (1997);  S. Dimopoulos, G. Dvali, R. Rattazzi, and G.F. Giudice,
    CERN-TH/97-98, hep-ph/9705307.
    
  \bibitem{HNY} J. Hisano, Y. Nomura, and T. Yanagida, KEK-TH-547,
    hep-ph/9710279.
    
  \bibitem{MK} H. Murayama and D. B. Kaplan, {\it Phys. Lett.}\/ {\bf
      B336}, 221 (1994).
    
  \bibitem{INS} H. Murayama, Invited talk presented at the 22nd INS
    International Symposium on Physics with High Energy Colliders,
    Tokyo, Japan, March 8--10, 1994, published in Proceedings of INS
    Symposium, World Scientific, 1994.
  
  \bibitem{Antaramian} A. Antaramian, LBL-36819, Feb 1995, Ph.D. Thesis.
    
  \bibitem{BB} K.S. Babu and S.M. Barr, \Journal{\PLB}{381}{137}{1996}.

  \bibitem{Rabyproton} V. Lucas and  S. Raby,
    \Journal{\PRD}{55}{6986}{1997}. 
    
  \bibitem{BPW} K.S. Babu, J. C. Pati, and F. Wilczek,
    IASSNS-HEP-97-136, hep-ph/9712307. 
      
  \bibitem{MNTY} A. Masiero, D.V. Nanopoulos, K. Tamvakis, and T.
    Yanagida, \Journal{\PLB}{115}{380}{1982}; B. Grinstein,
    \Journal{\NPB}{206}{387}{1982}.
    
  \bibitem{DWS} S. Dimopoulos and F. Wilczek, NSF-ITP-82-07
    (unpublished); M. Srednicki, \Journal{\NPB}{202}{327}{1982}.
    
  \bibitem{Yamada} K. Hagiwara and Y. Yamada,
    \Journal{\PRL}{70}{709}{1993}; Y. Yamada,
    \Journal{\ZPC}{60}{83}{1993}.
    
  \bibitem{missingPQ} J. Hisano, T. Moroi, K. Tobe, and T. Yanagida,
    \Journal{\PLB}{342}{138}{1995}.
    
  \bibitem{BB0} K.S. Babu and S.M. Barr,
    \Journal{\PRD}{48}{5354}{1993}.
    
  \bibitem{ELN} J. Ellis, J. L. Lopez, and D.V. Nanopoulos,
    \Journal{\PLB}{371}{65}{1996}.
    
  \bibitem{discreteB} L. E. Ibanez and G. G. Ross, {\it Nucl. Phys.}\/
    {\bf B368}, 3 (1992).
  
  \bibitem{Nirproton} V. Ben-Hamo and Y. Nir,
    \Journal{\PLB}{339}{77}{1994}.

  \bibitem{S3}
    L. J. Hall and H. Murayama, {\it Phys. Rev. Lett.}\/ {\bf 75},
    3985 (1995). 

  \bibitem{S3q} C. D. Carone, L. J. Hall, and H. Murayama,
    {\it Phys. Rev.}\/ {\bf D53}, 6282 (1996).
    
  \bibitem{seesaw} T. Yanagida, in {\it Proceedings of Workshop on the
      Unified Theory and the Baryon Number in the Universe},\/
    Tsukuba, Japan, 1979, edited by A. Sawada
    and A. Sugamoto (KEK, Tsukuba, 1979), p. 95;
    M. Gell-Mann, P. Ramond and R. Slansky, in {\it Supergravity},\/
    proceedings of the Workshop, Stony Brook, New York, 1979, edited
    by P. Van Nieuwenhuizen and D.Z. Freedman (North-Holland,
    Amsterdam, 1979), p. 315.
    
  \bibitem{neutrino} Y. Grossman and Y. Nir,
    \Journal{\NPB}{448}{30}{1995}; M. Schmaltz,
    \Journal{\PRD}{52}{1643}{1995}; C. D. Carone and L. J. Hall,
    \Journal{\PRD}{56}{4198}{1997}; P. Binetruy, S. Lavignac, S.
    Petcov, and P. Ramond, \Journal{\NPB}{496}{3}{1997}.

  \bibitem{degenerate} M.~Dine, A.~Kagan, and R.~Leigh,
    \Journal{\PRD}{48}{4269}{1993} ; P.~Pouliot and N.~Seiberg,
    \Journal{\PLB}{318}{169}{1993}; D.B.~Kaplan and M.~Schmaltz,
    \Journal{\PRD}{49}{3741}{1994}; A.~Pomarol and D.~Tommasini,
    \Journal{\NPB}{466}{3}{1996}; R. Barbieri, G. Dvali, and
    L. J. Hall, \Journal{\PLB}{377}{76}{1996}; R. Barbieri,
    L. J. Hall, S. Raby, and A. Romanino,
    \Journal{\NPB}{493}{3}{1997}. 


  \bibitem{align} Y.~Nir and N.~Seiberg, \Journal{\PLB}{309}{337}{1993}.
  
  \bibitem{HKR} L. J. Hall, V. A. Kostelecky, and S. Raby,
    \Journal{\NPB}{267}{415}{1986}.

  \bibitem{BH1} R. Barbieri and L.J. Hall,
    \Journal{\PLB}{338}{212}{1994}. 
    
  \bibitem{muegamma} R. Barbieri, L.J. Hall, and A.  Strumia,
    \Journal{\NPB}{445}{219}{1995}; J. Hisano, T. Moroi, K. Tobe, M.
    Yamaguchi, and T. Yanagida, \Journal{\PLB}{357}{579}{1995}.

  \bibitem{BHforK} R. Barbieri, L.J. Hall, and 
    A. Strumia, \Journal{\NPB}{449}{437}{1995}.

  \bibitem{EDM} S. Dimopoulos and L.J. Hall,
    \Journal{\PLB}{344}{185}{1995}. 
    
  \bibitem{ACFH} N.  Arkani-Hamed, H.-C. Cheng, J. L. Feng, and L. J.
    Hall, \Journal{\PRL}{77}{1937}{1996}.
        
 

 \end{thebibliography}
\end{document}